\documentclass[runningheads]{llncs}

\usepackage[T1]{fontenc}

\usepackage[hyphens]{url} 
\usepackage[hidelinks]{hyperref}
\usepackage[hyphenbreaks]{breakurl} 
\usepackage[all]{hypcap}

\usepackage{amsmath}

\newcommand{\deadlock}{\mathbf{0}} 
\newcommand{\newchannel}[1]{\nu#1.\>}
\newcommand{\bridge}{\mathbin\rightarrow}
\newcommand{\distributor}{\mathbin\Rightarrow}
\newcommand{\unreliability}[1]{\text{\textup{\textcurrency}}^{#1}}
\newcommand{\loser}{\unreliability{?}}
\newcommand{\duplicator}{\unreliability{+}}
\newcommand{\duploser}{\unreliability{*}}

\usepackage{tikz}
\usetikzlibrary{petri}

\tikzset{
    communication net/.style={
        every place/.style={minimum size=4.5mm},
        every transition/.style={minimum size=3mm},
        local/.style={double}
    },
    symbol/.style={
        anchor=center
    }
}

\newcommand{\losing}{$?$}
\newcommand{\duplicating}{$+$}
\newcommand{\duplosing}{$*$}

\newcommand{\verticaldots}{%
    \makebox[0cm][l]{\raisebox{0.3cm}{$\cdot$}}%
    \makebox[0cm][l]{\raisebox{-0.3cm}{$\cdot$}}%
    $\cdot$%
}

\begin{document}

\title{Your Blockchain Needn't Care\\How the Message is Spread}
\author{
    Wolfgang Jeltsch\inst{1,3}\orcidID{0000-0002-8068-8401}
\and
    Javier D\'{i}az\inst{2,3}\orcidID{0000-0003-3588-7468}
}
\institute{
    Well-Typed, London, England\\
    \email{wolfgang@well-typed.com}
\and
    Atix Labs (a Globant Division), Buenos Aires, Argentina\\
    \email{javier.diaz@globant.com}
\and
    IO Global, Boulder, Colorado
}
\maketitle

\begin{abstract}

In a blockchain system, nodes regularly distribute data to other nodes.
The ideal perspective taken in the scientific literature is that data is
broadcast to all nodes directly, while in practice data is distributed
by repeated multicast. Since correctness and security typically have
been established for the ideal setting only, it is vital to show that
these properties carry over to real-world implementations. This can be
done by proving that the ideal and the real behavior are equivalent.

In the work described in this paper, we take an important step towards
such a proof by proving a simpler variant of the above equivalence
statement. The simplification is that we consider only a concrete pair
of network topologies, which nevertheless illustrates important
phenomena encountered with arbitrary topologies. For describing systems
that distribute data, we use a domain-specific language of processes
that is embedded in a general-purpose process calculus. This allows us
to leverage the rich theory of process calculi in our proof, which is
machine-checked using the Isabelle proof assistant.


\end{abstract}

\section{Introduction}

Blockchains are becoming increasingly relevant in a variety of fields,
such as finance, identification, logistics, and real estate. Incorrect
behavior of a blockchain system may often result in serious damage.
Therefore, machine-checked proofs of correctness and security are of
high value in the blockchain field.

The fundamental task of a blockchain system is to maintain data
consistency among distributed agents in an open network. To facilitate
this, nodes have to regularly distribute data to other nodes. The
perspective usually taken in the scientific literature is that such data
is broadcast to all nodes directly. For example, the descriptions of the
blockchain consensus protocols of the Ouroboros
family~\cite{badertscher:2018,david:2018,kiayias:2017} assume such
direct communication. In practice, however, data is distributed via
repeated multicast. Since correctness and security typically have been
established for the ideal setting only, it is vital to show that they
carry over to real-world implementations.

In this paper, we take an important step in this direction by showing
that the ideal behavior of direct broadcast and the real behavior of
broadcast via multicast are equivalent in an appropriate sense.
Concretely, we make the following contributions:
\begin{itemize}

\item

In Sect.~\ref{section:communication-language}, we define a restricted
language of processes that are able to describe network communication.
Processes in our language are closely connected to hierarchical Petri
nets with exactly one input place per transition. We define our language
via an embedding in a general-purpose process calculus. This approach
enables us to leverage the rich theory of process calculi while allowing
us to use an intuitive graphical notation similar to the one of Petri
nets.

\item

In Sect.~\ref{section:behavioral-equivalence}, we devise a notion of
behavioral equivalence of networks that does not distinguish between
different patterns of packet arrival. Our approach is to start with
bisimilarity and weaken it by amending the involved processes to allow
for additional behavior. Building on bisimilarity permits us in
particular to reason in a modular fashion.

\item

In Sect.~\ref{section:correctness-proof}, we present a proof of
behavioral equivalence of broadcast via repeated multicast and direct
broadcast under the assumption that network communication may involve
packet loss and duplication. Our proof is about a concrete pair of
networks that nevertheless captures important general phenomena. The
proof works by rewriting a process describing the former form of
broadcast into a process describing the latter. For the individual
rewriting steps, we rely on certain fundamental lemmas, not all of which
have been proved so far. We present only the first part of our proof in
detail, using the graphical notation for processes, but we have
formalized the whole proof in Isabelle/HOL.

\end{itemize}
Afterwards, we discuss related work in Sect.~\ref{section:related-work}
and give a conclusion and an outlook on ongoing and future work in
Sects.~\ref{section:conclusion} and
\ref{section:ongoing-and-future-work}.

\section{A Language for Communication Networks}

\label{section:communication-language}

For describing communication networks, we use a custom language of
processes that communicate via asynchronous channels. Let uppercase
letters denote processes and lowercase letters denote channels. The
syntax of our communication language is given by the following BNF rule:
\begin{equation*}
\mathit{Process} \mathrel{{\mathop:}{\mathop:}{=}}
    \deadlock                             \mid
    P \parallel Q                         \mid
    \newchannel{a} P                      \mid
    a \bridge b                           \mid
    a \distributor [b_{1}, \ldots, b_{n}] \mid
    \loser a                              \mid
    \duplicator a                         \mid
    \duploser a
\end{equation*}
A process is one of the following:
\begin{itemize}

\item

The \emph{stop process}~$\deadlock$, which does nothing

\item

A \emph{parallel composition} $P \parallel Q$, which performs
$P$~and~$Q$ in parallel

\item

A \emph{restricted process} $\newchannel{a} P$, which behaves like~$P$
except that the channel~$a$ is local

\item

A \emph{bridge} $a \bridge b$, which continuously forwards packets from
channel~$a$ to channel~$b$

\item

A \emph{distributor} $a \distributor [b_{1}, \ldots, b_{n}]$, which
continuously forwards packets from channel~$a$ to all channels~$b_{i}$

\item

A \emph{loser} $\loser a$, which continuously drops packets from
channel~$a$

\item

A \emph{duplicator} $\duplicator a$, which continuously duplicates
packets in channel~$a$

\item

A \emph{duploser} $\duploser a$, which continuously drops packets from
and continuously duplicates packets in channel~$a$

\end{itemize}

Our language is embedded in the \TH-calculus\footnote{See
\url{https://github.com/input-output-hk/thorn-calculus}.}, a
general-purpose process calculus that is itself embedded in
Isabelle/HOL. The first three constructs of our language stem directly
from the \TH-calculus, distributors are defined in terms of the more
primitive send and receive constructs the \TH-calculus provides, and the
remaining constructs are defined in terms of the former ones.

\begin{figure}

\centering
\begin{tikzpicture}[communication net,caption/.style={anchor=base}]

\begin{scope}[example channel/.style={place,label=right:$a$}]

\node [example channel] at (-4,0)         {};
\node [example channel] at (-2,0) [local] {};
\node [example channel] at ( 0,0)         {\losing};
\node [example channel] at ( 2,0)         {\duplicating};
\node [example channel] at ( 4,0)         {\duplosing};

\end{scope}

\begin{scope}[yshift=-0.8cm]

\node [caption] at (-4,0) {$a$};
\node [caption] at (-2,0) {$\newchannel{a} \ldots a \ldots$};
\node [caption] at ( 0,0) {$\loser a$};
\node [caption] at ( 2,0) {$\duplicator a$};
\node [caption] at ( 4,0) {$\duploser a$};

\end{scope}

\begin{scope}[yshift=-3.5cm]

\begin{scope}

\node [place] (a) at (-3,1) [label=left:$a$]  {};
\node [place] (b) at (-1,1) [label=right:$b$] {};

\node [transition] at (-2,1) {} edge [pre]  (a)
                                edge [post] (b);

\end{scope}

\begin{scope}

\node [place] (a)  at (1,1) [label=left:$a$]      {};
\node [place] (b1) at (3,2) [label=right:$b_{1}$] {};
\node [place] (bn) at (3,0) [label=right:$b_{n}$] {};

\node [symbol] at (3, 1) {\verticaldots};

\node [transition] at (2,1) {} edge [pre]  (a)
                               edge [post] (b1)
                               edge [post] (bn);

\end{scope}

\begin{scope}[yshift=-0.8cm]

\node [caption] at (-2,0) {$a \bridge b$};
\node [caption] at ( 2,0) {$a \distributor [b_{1}, \ldots, b_{n}]$};

\end{scope}

\end{scope}

\end{tikzpicture}

\caption{Elements of communication nets}

\label{figure:communication-nets}

\end{figure}
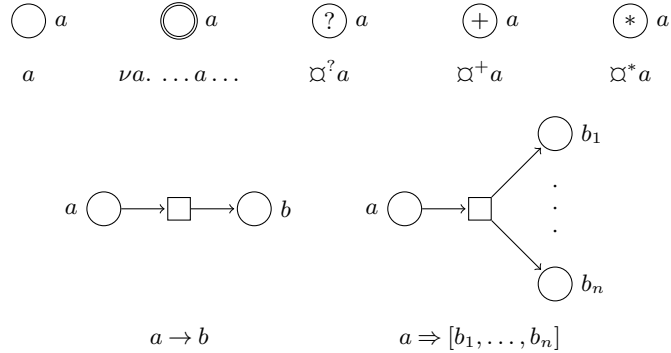

The sublanguage formed by $\deadlock$, $\parallel$, $\nu$, and
$\distributor$ corresponds closely to hierarchical Petri nets with
exactly one input place per transition, and the other constructs can be
analogously derived for such Petri nets. Based on this close connection
to Petri nets, we introduce a graphical representation of communication
processes as \emph{communication nets}, which is shown in
Fig.~\ref{figure:communication-nets}. Note that
$\deadlock$~and~$\parallel$ are represented by absence and composition,
respectively.

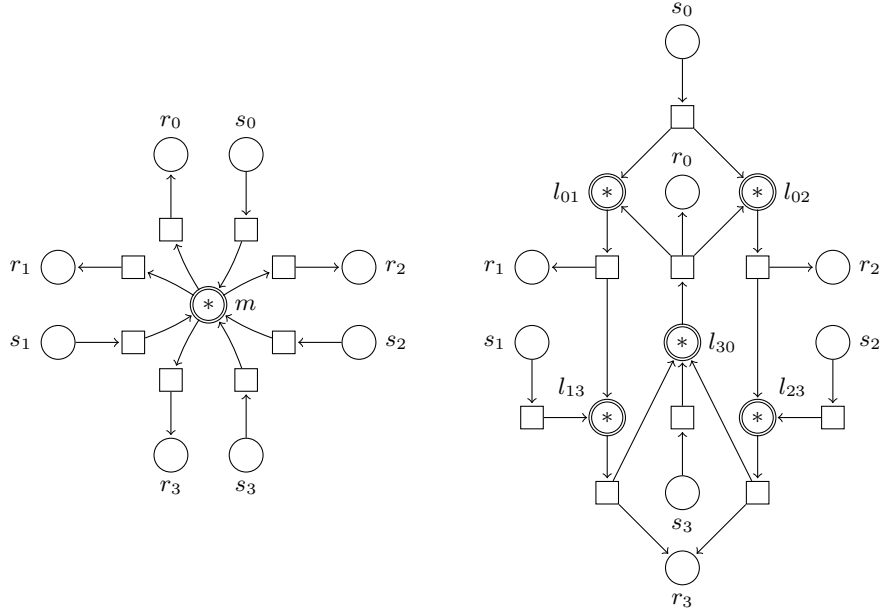
\begin{figure}

\centering
\begin{tikzpicture}[communication net]

\begin{scope}[bend angle=6]

\node [place] (s0) at ( 0.5,   2) [label=above:$s_{0}$]   {};
\node [place] (s1) at (  -2,-0.5) [label=left:$s_{1}$]    {};
\node [place] (s2) at (   2,-0.5) [label=right:$s_{2}$]   {};
\node [place] (s3) at ( 0.5,  -2) [label=below:$s_{3}$]   {};
\node [place] (r0) at (-0.5,   2) [label=above:$r_{0}$]   {};
\node [place] (r1) at (  -2, 0.5) [label=left:$r_{1}$]    {};
\node [place] (r2) at (   2, 0.5) [label=right:$r_{2}$]   {};
\node [place] (r3) at (-0.5,  -2) [label=below:$r_{3}$]   {};
\node [place] (m)  at (   0,   0) [local,label=right:$m$] {\duplosing};

\node [transition] at ( 0.5,   1) {} edge [pre]             (s0)
                                     edge [post,bend left]  (m);
\node [transition] at (  -1,-0.5) {} edge [pre]             (s1)
                                     edge [post,bend right] (m);
\node [transition] at (   1,-0.5) {} edge [pre]             (s2)
                                     edge [post,bend left]  (m);
\node [transition] at ( 0.5,  -1) {} edge [pre]             (s3)
                                     edge [post,bend right] (m);
\node [transition] at (-0.5,   1) {} edge [pre,bend right]  (m)
                                     edge [post]            (r0);
\node [transition] at (  -1, 0.5) {} edge [pre,bend left]   (m)
                                     edge [post]            (r1);
\node [transition] at (   1, 0.5) {} edge [pre,bend right]  (m)
                                     edge [post]            (r2);
\node [transition] at (-0.5,  -1) {} edge [pre,bend left]   (m)
                                     edge [post]            (r3);

\end{scope}

\begin{scope}[xshift=6.3cm,yshift=-0.5cm]

\newcommand{\dl}{\duplosing}

\node [place] (s0)  at ( 0, 4) [label=above:$s_{0}$]              {};
\node [place] (s1)  at (-2, 0) [label=left:$s_{1}$]               {};
\node [place] (s2)  at ( 2, 0) [label=right:$s_{2}$]              {};
\node [place] (s3)  at ( 0,-2) [label=below:$s_{3}$]              {};
\node [place] (r0)  at ( 0, 2) [label=above:$r_{0}$]              {};
\node [place] (r1)  at (-2, 1) [label=left:$r_{1}$]               {};
\node [place] (r2)  at ( 2, 1) [label=right:$r_{2}$]              {};
\node [place] (r3)  at ( 0,-3) [label=below:$r_{3}$]              {};
\node [place] (l01) at (-1, 2) [local,label=left:$l_{01}$]        {\dl};
\node [place] (l02) at ( 1, 2) [local,label=right:$l_{02}$]       {\dl};
\node [place] (l13) at (-1,-1) [local,label=above left:$l_{13}$]  {\dl};
\node [place] (l23) at ( 1,-1) [local,label=above right:$l_{23}$] {\dl};
\node [place] (l30) at ( 0, 0) [local,label=right:$l_{30}$]       {\dl};

\node [transition] at ( 0, 3) {} edge [pre]  (s0)
                                 edge [post] (l01)
                                 edge [post] (l02);
\node [transition] at (-2,-1) {} edge [pre]  (s1)
                                 edge [post] (l13);
\node [transition] at ( 2,-1) {} edge [pre]  (s2)
                                 edge [post] (l23);
\node [transition] at ( 0,-1) {} edge [pre]  (s3)
                                 edge [post] (l30);
\node [transition] at (-1, 1) {} edge [pre]  (l01)
                                 edge [post] (r1)
                                 edge [post] (l13);
\node [transition] at ( 1, 1) {} edge [pre]  (l02)
                                 edge [post] (r2)
                                 edge [post] (l23);
\node [transition] at (-1,-2) {} edge [pre]  (l13)
                                 edge [post] (r3)
                                 edge [post] (l30);
\node [transition] at ( 1,-2) {} edge [pre]  (l23)
                                 edge [post] (r3)
                                 edge [post] (l30);
\node [transition] at ( 0, 1) {} edge [pre]  (l30)
                                 edge [post] (r0)
                                 edge [post] (l01)
                                 edge [post] (l02);

\end{scope}

\end{tikzpicture}

\caption{%
    Example of direct broadcast (left)
    and broadcast via multicast (right)%
}

\label{figure:broadcast}

\end{figure}

Two complete communication nets are shown in
Fig.~\ref{figure:broadcast}. They characterize the two networks whose
behavioral equivalence we will show in
Sect.~\ref{section:correctness-proof}. In both communication nets,
channels $s_{i}$ and $r_{i}$ form the interface of a network node~$i$,
with $s_{i}$ accepting packets for sending and $r_{i}$ providing
received packets. The local channel~$m$ in the left communication net
represents the broadcast medium, and each local channel $l_{ij}$ in the
right communication net represents a multicast link from node~$i$ to
node~$j$. Note that we assume any communication to be unreliable, which
is reflected by the channel~$m$ and all channels~$l_{ij}$ having
duplosers attached.

The two communication nets correspond to processes~$D$ (direct
broadcast) and~$M$ (broadcast via multicast) defined as follows:
\begin{align}
\label{equation:direct-broadcast}
D & =
\newchannel{m}
(
    \duploser m
    \parallel s_{0} \bridge m \parallel \ldots \parallel s_{3} \bridge m
    \parallel m \bridge r_{0} \parallel \ldots \parallel m \bridge r_{3}
)
\\
\label{equation:broadcast-via-multicast}
M & =
\newchannel{l_{01}}
\newchannel{l_{02}}
\newchannel{l_{13}}
\newchannel{l_{23}}
\newchannel{l_{30}}
(M_{*} \parallel M_{\mathrm{i}} \parallel M_{\mathrm{o}})
\\
M_{*} & =
\duploser l_{01} \parallel
\duploser l_{02} \parallel
\duploser l_{13} \parallel
\duploser l_{23} \parallel
\duploser l_{30}
\\
M_{\mathrm{i}} & =
s_{0} \distributor [l_{01}, l_{02}] \parallel
s_{1} \distributor [l_{13}]         \parallel
s_{2} \distributor [l_{23}]         \parallel
s_{3} \distributor [l_{30}]
\\
M_{\mathrm{o}} & =
l_{01} \distributor [r_{1}, l_{13}]         \parallel
l_{02} \distributor [r_{2}, l_{23}]         \parallel
l_{13} \distributor [r_{3}, l_{30}]         \parallel
l_{23} \distributor [r_{3}, l_{30}]         \parallel
\notag \\ & \mathrel{\phantom{=}}
l_{30} \distributor [r_{0}, l_{01}, l_{02}]
\end{align}

\section{Loss-Agnostic Behavioral Equivalence}

\label{section:behavioral-equivalence}

Weak bisimilarity would be a natural choice for the kind of equivalence
that should hold between the two network processes shown in the previous
section. Weak bisimilarity is a well-established notion of behavioral
equivalence that provides a fine-grained distinction of observable
behavior and allows for modular reasoning. For our communication
language, its notion of behavior characterizes essentially how handing
over packets to global channels may result in packets appearing on
possibly other global channels.

Unfortunately, weak bisimilarity turns out to be too strict for our
situation, as it is able to distinguish between broadcast via multicast
and direct broadcast. To see why, assume in each of the networks shown
in Fig.~\ref{figure:broadcast} a packet is sent by node~$0$ and this
packet makes it to node~$3$. With direct broadcast, it is possible that
neither node~$1$ nor node~$2$ receives the packet as well. With
broadcast via multicast, however, node~$1$ or node~$2$ must receive it
and must do so before node~$3$ receives it.

To remove this constraint on arrival patterns, we make the receive
channels lossy. This way, intermediate nodes are no longer guaranteed to
receive packets. It is not sufficient, however, to introduce this
lossiness for broadcast via multicast only; we need to introduce it also
for direct broadcast. This is because packet loss in the receive
channels is observable and consequently unilateral introduction of such
loss would create another behavioral mismatch between the two networks.
The approach of making receive channels lossy leads to a notion of
\emph{weak bisimilarity up to loss}, which is derived from weak
bisimilarity~(written $\approx$ in the following):
\begin{definition}[Weak bisimilarity up to loss]
Two processes $P$ and~$Q$ are weakly bisimilar up to loss in channels
$r_{1}$ to~$r_{n}$ exactly if
\begin{equation*}
\loser r_{1} \parallel \ldots \parallel \loser r_{n} \parallel P
\approx
\loser r_{1} \parallel \ldots \parallel \loser r_{n} \parallel Q
\enspace.
\end{equation*}
\end{definition}

\section{A Proof of Correctness of Broadcast via Multicast}

\label{section:correctness-proof}

Broadcast via multicast is expected to behave equivalently to direct
broadcast as long as the multicast network is strongly connected. In
this work, however, we prove this equivalence only for the particular
networks depicted in Fig.~\ref{figure:broadcast}, which nevertheless
capture important phenomena that show up in other cases:
\begin{itemize}

\item

The multicast network has a node with several outgoing and a node with
several incoming links.

\item

In the multicast network, some nodes are reachable from certain other
nodes only via more than one hop.

\end{itemize}

Concretely, our goal is to prove that $M$, defined in
Eq.~\ref{equation:broadcast-via-multicast}, and $D$, defined in
Eq.~\ref{equation:direct-broadcast}, are weakly bisimilar up to loss in
the receive channels, which is expressed by the statement
\begin{equation*}
\loser r_{0} \parallel \ldots \parallel \loser r_{3} \parallel M
\approx
\loser r_{0} \parallel \ldots \parallel \loser r_{3} \parallel D
\enspace.
\end{equation*}
We prove this statement by turning its left-hand side into its
right-hand side through a series of transformation steps, each of which
replaces subprocesses with bisimilar ones. The individual transformation
steps rely on several fundamental lemmas about the communication
language, not all of which we have proved yet.

The first transformation step deals with the distributors that forward
packets from link channels. These distributors deliver packets from the
link channels~$l_{ij}$ to the receive channels~$r_{j}$ and also relay
them to the follow-up link channels~$l_{jk}$. The first transformation
step splits each of these distributors into multiple bridges and thus in
particular separates delivery and relaying of packets. The subsequent
transformation steps collapse the relaying part into a single channel:
the broadcast medium. We only discuss the first step in detail.

The first transformation step takes the broadcast-via-multicast process
depicted in Fig.~\ref{figure:broadcast} with the receive channels made
lossy and turns it into the process depicted in
Fig.~\ref{figure:after-distributor-splitting}. The key justification for
this transformation is the existence of the \emph{distributor splitting
lemma}, which is shown in Fig.~\ref{figure:distributor-splitting}. Note
that, in order to apply this lemma, the respective source channel must
have an accompanying duplicator and the respective target channels must
have accompanying losers. The link channels, which also act as target
channels and additionally as source channels, have duplosers attached to
them. Since a duploser is defined as a parallel composition of a loser
and a duplicator, the link channels fulfill the conditions we require
for source and target channels.
\begin{figure}

\centering
\begin{tikzpicture}[communication net]

\newcommand{\ls}{\losing}
\newcommand{\dl}{\duplosing}

\node [place] (s0)  at ( 0, 4) [label=above:$s_{0}$]        {};
\node [place] (s1)  at (-4, 0) [label=left:$s_{1}$]         {};
\node [place] (s2)  at ( 4, 0) [label=right:$s_{2}$]        {};
\node [place] (s3)  at ( 0,-2) [label=below:$s_{3}$]        {};
\node [place] (r0)  at ( 0, 2) [label=above:$r_{0}$]        {\ls};
\node [place] (r1)  at (-4, 1) [label=left:$r_{1}$]         {\ls};
\node [place] (r2)  at ( 4, 1) [label=right:$r_{2}$]        {\ls};
\node [place] (r3)  at ( 0,-3) [label=below:$r_{3}$]        {\ls};
\node [place] (l01) at (-2, 2) [local,label=left:$l_{01}$]  {\dl};
\node [place] (l02) at ( 2, 2) [local,label=right:$l_{02}$] {\dl};
\node [place] (l13) at (-2,-1) [local,label=left:$l_{13}$]  {\dl};
\node [place] (l23) at ( 2,-1) [local,label=right:$l_{23}$] {\dl};
\node [place] (l30) at ( 0, 0) [local,label=right:$l_{30}$] {\dl};

\node [transition] at ( 0,  3) {} edge [pre]  (s0)
                                  edge [post] (l01)
                                  edge [post] (l02);
\node [transition] at (-3,  0) {} edge [pre]  (s1)
                                  edge [post] (l13);
\node [transition] at ( 3,  0) {} edge [pre]  (s2)
                                  edge [post] (l23);
\node [transition] at ( 0, -1) {} edge [pre]  (s3)
                                  edge [post] (l30);
\node [transition] at (-3,  1) {} edge [pre]  (l01)
                                  edge [post] (r1);
\node [transition] at ( 3,  1) {} edge [pre]  (l02)
                                  edge [post] (r2);
\node [transition] at (-2, -2) {} edge [pre]  (l13)
                                  edge [post] (r3);
\node [transition] at ( 2, -2) {} edge [pre]  (l23)
                                  edge [post] (r3);
\node [transition] at ( 0,  1) {} edge [pre]  (l30)
                                  edge [post] (r0);
\node [transition] at (-2,0.5) {} edge [pre]  (l01)
                                  edge [post] (l13);
\node [transition] at ( 2,0.5) {} edge [pre]  (l02)
                                  edge [post] (l23);
\node [transition] at (-1, -1) {} edge [pre]  (l13)
                                  edge [post] (l30);
\node [transition] at ( 1, -1) {} edge [pre]  (l23)
                                  edge [post] (l30);
\node [transition] at (-1,  1) {} edge [pre]  (l30)
                                  edge [post] (l01);
\node [transition] at ( 1,  1) {} edge [pre]  (l30)
                                  edge [post] (l02);

\end{tikzpicture}

\caption{%
    The broadcast-via-multicast process
    after distributor splitting%
}

\label{figure:after-distributor-splitting}

\end{figure}
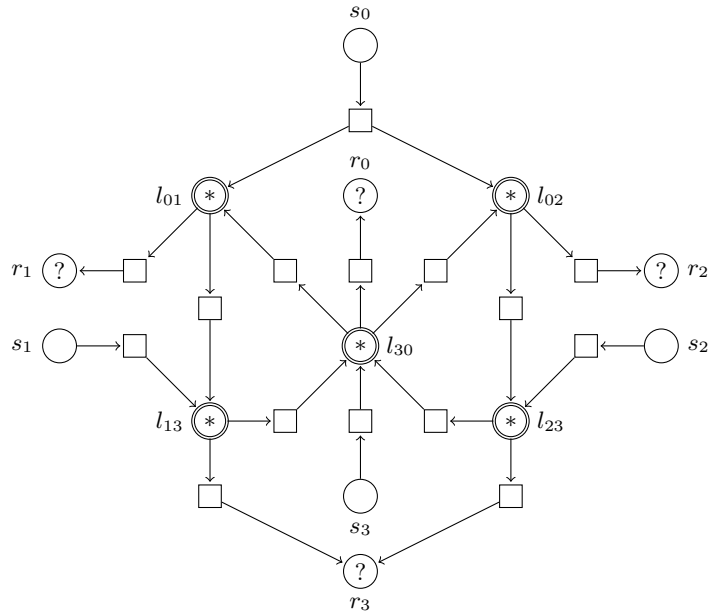
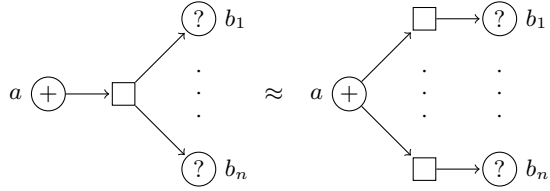
\begin{figure}

\centering
\begin{tikzpicture}[communication net]

\begin{scope}

\node [place] (a)  at (-3, 0) [label=left:$a$]      {\duplicating};
\node [place] (b1) at (-1, 1) [label=right:$b_{1}$] {\losing};
\node [place] (bn) at (-1,-1) [label=right:$b_{n}$] {\losing};

\node [symbol] at (-1, 0) {\verticaldots};

\node [transition] at (-2,0) {} edge [pre]  (a)
                                edge [post] (b1)
                                edge [post] (bn);

\end{scope}

\node [symbol] at (0,0) {$\approx$};

\begin{scope}

\node [place] (a)  at (1, 0) [label=left:$a$]      {\duplicating};
\node [place] (b1) at (3, 1) [label=right:$b_{1}$] {\losing};
\node [place] (bn) at (3,-1) [label=right:$b_{n}$] {\losing};

\node [symbol] at (3, 0) {\verticaldots};

\node [transition] at (2, 1) {} edge [pre]  (a)
                                edge [post] (b1);
\node [transition] at (2,-1) {} edge [pre]  (a)
                                edge [post] (bn);

\node [symbol] at (2, 0) {\verticaldots};

\end{scope}

\end{tikzpicture}

\caption{The distributor splitting lemma}

\label{figure:distributor-splitting}

\end{figure}

We have conducted the complete proof using communication nets and
additionally developed a formal version of it\footnote{See
\url{https://github.com/input-output-hk/network-equivalences}.} using
the Isabelle proof assistant. The formal version is phrased in a style
similar to equational reasoning, with the difference that we use weak
bisimilarity instead of equality. Since Isabelle does not come with
support for automated rewriting based on equivalence relations other
than equality, we have developed a corresponding extension\footnote{See
\url{https://github.com/input-output-hk/equivalence-reasoner}.}, which
we use in our proof.

While the formal proof closely follows its communication net
counterpart, it contains more technical details. In particular, it
includes applications of basic bisimilarities, like associativity and
commutativity of parallel composition, which the communication net proof
does not need, because the communication net notation identifies
processes that are bisimilar according to these basic bisimilarities.
However, all these bisimilarities together form a confluent and
terminating rewrite system, allowing our engine for equivalence-based
rewriting to automatically find any proof that consists of a chain of
rewriting steps involving only these bisimilarities. As a result, we can
bundle consecutive applications of said basic bisimilarities in our
formal proof, which therefore is still reasonably concise and readable,
while being machine-checked at the same time.

\section{Related Work}

\label{section:related-work}

The correctness of broadcast via multicast, commonly referred to as
\emph{network flooding}, and similar techniques has been studied to some
extent in the literature. Following are two examples:
\begin{itemize}

\item

Bani-Abdelrahman~\cite{bani-abdelrahman:2018} formally specifies
synchronous and bounded asynchronous flooding algorithms using LTL and
verifies them using the model checker NuSMV. His results are limited to
small network sizes and fixed delays, though.

\item

Bar-Yehuda et~al.~\cite{bar-yehuda:1989} emulate a single-hop
(direct-broadcast) network with a multi-hop network using a synchronous
gossiping algorithm. With a gossiping algorithm, a node does not relay
an incoming packet to all neighboring nodes, but only to a randomly
chosen one.

\end{itemize}
The existing literature, however, seems to lack the study of an
\emph{equivalence} of the two broadcasting approaches, which we provide
in this work. The reason for this lack may be related to the
complications that arise due to the behavioral mismatches explained in
Sect.~\ref{section:behavioral-equivalence}.

There is a vast amount of literature on the relationship between process
calculi and Petri nets. In particular, a long line of research has been
developed with the theme of giving Petri net semantics to process
calculi~\cite{best:2001,degano:1988,olderog:1991}, providing process
calculi with operational semantics expressing true concurrency as
opposed to the traditional interleaving semantics. Another line of
research approaches the reverse problem, that is, finding process
calculi that are suitable for modeling Petri nets of certain
classes~\cite{basten:1995,gorrieri:2017}. In the light of all this
research, our tandem of the communication language and its communication
net notation, described in Sect.~\ref{section:communication-language},
can be regarded as a restricted process calculus that comes with a Petri
net semantics.

\section{Conclusion}

\label{section:conclusion}

We have defined a language for describing communication networks, which
is embedded in a process calculus and closely connected to a class of
Petri nets. Based on the connection to Petri nets, we have devised a
graphical notation for our language. Building on this foundation, we
have proved behavioral equivalence of broadcast via multicast and direct
broadcast for a typical pair of networks. The graphical notation has
allowed us to reason in an intuitive way, while the embedding in a
process calculus has permitted us to develop a fully machine-checked
proof. For specifying the equivalence between the two realizations of
broadcast, we have devised the notion of weak bisimilarity up to loss.

\section{Ongoing and Future Work}

\label{section:ongoing-and-future-work}

At the moment, we are completing the proofs of the fundamental lemmas
that the correctness proof shown in
Sect.~\ref{section:correctness-proof} uses. Furthermore, we are working
on a variant of our correctness proof that deals with broadcast
integrated with packet filtering according to a fixed predicate.

In the future, we want to prove a modified correctness statement where
the receive channels of direct broadcast are not lossy. This will
clarify that broadcast via multicast has the more constrained behavior,
but will require adjustments to the specification of network behavior.
Furthermore, we want to generalize our correctness proofs to apply to
arbitrary strongly connected multicast networks and their
direct-broadcast counterparts. Finally, we want to generalize the proof
about broadcast integrated with filtering to work with state-dependent
filters.

\subsubsection{Acknowledgements}

We want to thank James Chapman, Duncan Coutts, Kevin Hammond, and
Philipp Kant, who supported us in our work on network equivalences and
the development of all the theory and the vast amount of Isabelle
formalizations that underlie it. We appreciate IO Global funding this
very interesting work.

\bibliographystyle{splncs04}
\bibliography{paper}

\end{document}